# Numerical analysis of the spin-orbit coupling parameters in III-V quantum wells using 8-band Kane model and finite-difference method

V E Degtyarev, S V Khazanova, A A Konakov[a)] and Yu A Danilov

Faculty of Physics, National Research Lobachevsky State University of Nizhny Novgorod, Nizhny Novgorod, Russia

[a)] E-mail: anton.a.konakov@gmail.com

**Abstract**

By means the envelope function approximation, 8-band Kane model and a finite-difference scheme with the coordinate space discretization, we numerically performed calculations of the spin-orbit coupling parameters for 2D electron gas confined in both symmetric and asymmetric [0 0 1] quantum wells based on zinc-blende III-V semiconductors. Influence of the quantum well band parameters and width as well as the magnitude of the external electric field applied along the growth direction on the Dresselhaus and Rashba spin-orbit coupling parameters is investigated. It has been found that in the symmetric InGaAs/GaAs quantum wells linear-in-momentum spin-orbit coupling disappears for the third electron subband at certain values of well width and the indium content. It is also shown that in asymmetric InGaAs/GaAs structures the spin-orbit coupling parameters can be equal at a certain electric field that is the condition for the realization of the SU(2) spin symmetry and formation of persistent spin helices. Besides, we calculated the spin-orbit coupling in the persistent spin helix regime as a function of the well width, indium content and external field. The proposed approach for the calculation of the spin-orbit coupling parameters can be applied to other 2D structures with the spin-orbit coupling.

## 1. Introduction

Due to effects of quantum confinement a magnitude of spin-dependent phenomena sufficiently increases in low-dimensional semiconductor structures [1]. Thus, two-dimensional (2D) electron systems are actively investigated for design of current and future spintronic devices [2]. A principal interest to 2D structures in the spintronics applications is determined by the spin-orbit coupling (SOC) that leads to the spin splitting and subsequent reconstruction of their energy spectrum [3] and causes a number of physical phenomena, such as spin-galvanic effect [4], spin Hall effect [5] and spin-dependent tunneling [6, 7], which are key for creation of devices based on spin-dependent phenomena. Presence of heavy elements (indium, antimony, bismuth) in heterostructures and influence of external potential can lead to even more noticeable spin-related phenomena and, consequently, need for detailed study of SOC.

In 2D crystalline systems there are two main linear-in-wavevector (linear-in-$k$) contributions to SOC: the first one is the Dresselhaus SOC due to asymmetry of a host crystal unit cell [8] and the second one is the Rashba interaction caused by asymmetry of a heterostructure [9]. The latter one can be effectively controlled by an applied bias [10–12]: an external electric field modifies the Rashba SOC parameter and, therefore, significantly affects kinetic [13, 14] and dynamical [15, 16] spin characteristics of a 2D electron gas. Moreover, Rashba and Dresselhaus terms can interfere resulting in the spin splitting anisotropy. In particular, for electron states in III-V quantum wells (QWs) with the [0 0 1] growth direction the SU(2) spin symmetry and corresponding persistent spin helix (PSH) state can be realized at the certain SOC parameters relations [17, 18]. Currently physics and applications of the PSH regime in 2D systems are of a great interest [19, 20, 21] due to the suppression of the D'yakonov-Perel' electron spin relaxation mechanism [22] and reduction of the spin relaxation rate [23]. Therefore, manipulation of the SOC parameters and PSH state by means an external electric field extends design possibilities of spintronic devices.

It should be noted that experimental determination of the SOC parameters remains a difficult task [24]. In this regard, their preliminary calculation allows one to predict spin-related properties of low-dimensional structures, compare them with experimental data and accordingly design QWs with required features.

Earlier the spin splitting of electron states in direct-band III-V QWs were calculated in works of Pfeffer and Zawadzki [25, 26] and Pfeffer [27, 28]. In Refs. [25, 27, 28] the envelope function approximation with the 14-band Kane model was used in order to receive analytical expressions for Rashba and Dresselhaus spin splitting, while in [26] the SOC parameters were calculated with the "conventional" 8-band Kane model. All calculations were performed taking into account the discontinuity of band parameters and external electric field, generating the asymmetry of the QW. The same multiband envelope function approximation based on both $8\times8$ and $14\times14$ Hamiltonians was used in [29]. One of the main conclusions of this work that application of 14-band Kane model to calculations of the spin splitting gives the minor improvements over results obtained with the use of "conventional" $8\times8$ model, while the number of parameters, equations and, in this way, complexity of calculations rises sufficiently. Thus, the use of 8-band Kane model adopted for calculations of spin-dependent properties is more preferable.

While above mentioned works concentrate on the SOC parameters calculation for the lowest lying electron subband, some experimental data show, that in the conduction band of III-V QWs there can be occupied more, than one subband [30–33]. For such systems it is important to know the SOC parameters for several quantum subbands. Theoretical description of the SOC in QWs with two subbands was first proposed in [34] and developed then in [35, 36]. In these works authors using 8-band Kane model received the effective 2-subband Hamiltonian in the envelope's subspace containing both intra- and intersubband linear-in-$k$ SOC. Note that for low energies and long wavelengths the correction to the spectra from conventional intrasubband SOC is linear-in-$k$, while contribution to the dispersion curves from intersubband SOC starts from the second order in the wave vector [35].

In all of these studies [25–28, 34–36] there were used, in fact, the averaging the additional SOC (Rashba or Rashba and Dresselhaus) over the states, obtained in the envelope function approximation, that was first proposed by D'yakonov and Kachorovsky in 1986 [37]. The main advantage of this approach is that it is semi-analytical way to obtain the expressions for the SOC parameters: it is necessary to calculate the envelope function and then determine essential matrix elements. However, here we show that it is only the first approximation to the SOC parameters, because some information about spectra and wave functions is lost through the averaging procedure.

In this work, we use 8-band Kane model and direct numerical approach to simulate the SOC parameters that enables increasing calculation accuracy and taking into account band structure peculiarities. Computer simulation and numerical calculations allow one to reveal the correlation between the technological parameters of growth and the SOC magnitudes in heterostructures with an arbitrary composition and geometry. Our aim is to define the spin splitting for different energy subbands and to investigate the possibility of controlling the SOC parameters for various QWs.

In particular, we propose a technique to extract the Dresselhaus parameter for symmetric heterostructures and quantitatively determine the Rashba interaction in the presence of the structure asymmetry caused by the electric bias applied along the growth direction. Moreover, we search for a tendency, how the wave function penetration in the barrier region of QW changes the SOC parameters. We also reveal factors that affect the ratio between Rashba and Dresselhaus contributions, which is important for realization of the PSH regime. All our calculations are carried out for In$_x$Ga$_{1-x}$As/GaAs QWs.

This paper is organized as following. In Sec. 2 we present theoretical background of the SOC parameters definition in zinc-blende semiconductor 2D systems and propose new numerical approach for their calculation. In Sec. 3 we demonstrate and discuss calculated Rashba and Dresselhaus parameters for In$_x$Ga$_{1-x}$As/GaAs QWs as functions of QW width, indium content and external bias. Peculiar attention is devoted to parameters that allow realization of the PSH state. Finally, Sec. 4 conclude the work.

## 2. Theoretical background and method of calculations

*2.1 The spin-orbit coupling in zinc-blende type quantum wells*

In this work we theoretically investigated the electronic states in In$_x$Ga$_{1-x}$As/GaAs QWs with account of SOC. As known, the effective Hamiltonian near the bottom of the conduction band of a bulk crystal with a zinc blende structure up to terms of the third order in the wave vector has the form [8]:

$$\hat{H} = \frac{\hbar^2 \mathbf{k}^2}{2m}\hat{\sigma}_0 + \gamma_D(\kappa_x \hat{\sigma}_x + \kappa_y \hat{\sigma}_y + \kappa_z \hat{\sigma}_z), \quad (1)$$

where $m$ is electron effective mass, $\kappa_x = k_x(k_y^2 - k_z^2)$, $\kappa_y = k_y(k_z^2 - k_x^2)$ and $\kappa_z = k_z(k_x^2 - k_y^2)$ are combinations of wave vector components due to effective magnetic field, caused by SOC, $k_x$, $k_y$ and $k_z$ are three components of the wave vector, $\hat{\sigma}_x$, $\hat{\sigma}_y$, $\hat{\sigma}_z$ and $\hat{\sigma}_0$ are Pauli matrices and unit matrix respectively, $\gamma_D$ is bulk Dresselhaus constant. In other words, the second term of (1) is bulk Dresselhaus SOC caused by the absence of an inversion centre in the crystal unit cell. Furthermore in the work, for clarity, the term "SOC parameter" refers to 2D system, while "SOC constant" corresponds to bulk Dresselhaus SOC. The SOC 3D Hamiltonian can be rewriten in explicit form as

$$\hat{H}_{3D} = \gamma_D k_z \hat{\sigma}_z (k_x^2 - k_y^2) - \gamma_D k_z^2 (k_x \hat{\sigma}_x - k_y \hat{\sigma}_y) + \\ + \gamma_D (k_x k_y^2 \hat{\sigma}_x - k_y k_x^2 \hat{\sigma}_y) \quad (2)$$

Confinement of carriers in 2D structures results into modification of SOC. Assuming that spectrum is quantized in the QW growth direction, for small 2D wave vectors electron states are described by two-component (taking spin into account) envelope function $\psi_{n,s}$, where $n$ is QW subband index and $s$ characterizes the spin projection. We suppose that QWs are grown in $[001]$ crystallographic direction, which corresponds to the $z$-axis in Hamiltonian (2).

Difference between material parameters in QW structure leads to need of the symmetrization procedure for $k$-dependent terms in the growth direction [38]:

$$k_z \to -i\frac{\partial}{\partial z}, \\ \gamma_D k_z \to -\frac{i}{2}\left(\frac{\partial}{\partial z}\gamma_D(z) + \gamma_D(z)\frac{\partial}{\partial z}\right), \quad (3) \\ -\gamma_D k_z^2 \to \frac{\partial}{\partial z}\gamma_D(z)\frac{\partial}{\partial z}.$$

After averaging over the $n^{th}$ QW subband with envelope function $\psi_{n,s}(z)$ the Hamiltonian (2) takes the form [26, 37]:

$$\hat{H}_{2D}^{(n)} = \beta_n(k_x\hat{\sigma}_x - k_y\hat{\sigma}_y) + \gamma_D^{(n)}(k_xk_y^2\hat{\sigma}_x - k_yk_x^2\hat{\sigma}_y), \quad (4)$$

where

$$\beta_n = \left\langle \frac{\partial}{\partial z}\gamma_D(z)\frac{\partial}{\partial z}\right\rangle_n \quad (5)$$

is the Dresselhaus parameter in the $n^{th}$ subband, $\gamma_D^{(n)} = \langle \gamma_D(z)\rangle_n$, and brackets $\langle ...\rangle_n$ stands for averaging over envelope function in the $n^{th}$ subband.

Asymmetry of the potential profile results in addition of linear in wave vector spin-orbit contribution to the Hamiltonian (the so-called Rashba term) [9]:

$$\hat{H}_R^{(n)} = \alpha_n(k_y\hat{\sigma}_x - k_x\hat{\sigma}_y). \quad (5)$$

where $\alpha_n$ is Rashba parameter in the $n^{th}$ subband. The asymmetry can be implemented with external electric field applied along the strcuture growth direction. In this case the Rashba parameter for each subband is proportional to the magnitude of applied field [9].

In QW structures the SOC contribution into electron Hamiltonian should contain both linear and cubic in wave vector terms. Excluding third-order wave vector terms in Hamiltonian (4) [26] and the interface contributions [39, 40], the effective 2D Hamiltonian for electrons in $n^{th}$ subband in zinc blende semiconductor asymmetric QW can be written as

$$\hat{H}_{2D} = \frac{\hbar^2(k_x^2 + k_y^2)}{2m}\hat{\sigma}_0 + \\ + \alpha_n(k_y\hat{\sigma}_x - k_x\hat{\sigma}_y) + \beta_n(k_x\hat{\sigma}_x - k_y\hat{\sigma}_y) \quad (6)$$

It should be noted that the Hamiltonian (6) is applicable only for small wave vectors near the QW conduction band bottom, therefore it describes the dispersion only for the first electron subband [41]. Thus, more accurate calculation of the QW energy spectra and SOC parameters requires usage of multi-band $\mathbf{k}\cdot\mathbf{p}$-models [42, 43], specifically, the 8-band Kane model [44, 45]. Within the framework of this approach, the $\mathbf{k}\cdot\mathbf{p}$-interaction of conduction and valence bands is taken into account exactly, while influence of all other bands is treated as a perturbation. Using this method the heterostructure energy spectrum and corresponding 8-component envelope function $\Psi_{n,s}(k_x, k_y, z)$ are calculated from Schrödinger-like equation

$$\left(\hat{H}_{8\times8}\left(k_x, k_y, -i\frac{\partial}{\partial z}\right) + \hat{I}_{8\times8}V(z)\right)\Psi_{n,s}(k_x, k_y, z) = \\ = E_{n,s}(k_x, k_y)\Psi_{n,s}(k_x, k_y, z), \quad (7)$$

where $\hat{H}_{8\times 8}(k_x, k_y, k_z)$ is the Kane Hamiltonian [44], $V(z)$ contains the confining potential including the asymmetry induced by external gate, $E_{n,s}(k_x, k_y)$ is the energy of the electronic state in $n^{th}$ subband with 2D wave vector **k** and the spin projection $s$.

*2.2. Numerical method of calculations*

In general, the analytical solution of equation (7) is difficult and the most universal approach is based on the numerical methods. For numerical solution of equation (7) we used the finite difference method (FDM). In the framework of this method Hamiltonian parameters (let us denote them as $C(z)$) and envelope functions are determined as spatial functions on the nodes of a one-dimensional uniform grid in the QW growth direction:

$$C(z) \to C(z_0 + jdz) \equiv C_j,$$
$$\Psi(k_x, k_y, z) \to \Psi(k_x, k_y, z_0 + jdz) = \Psi_j, \quad (8)$$

where $dz$ is the grid spacing. In order to get the solution we convert the differential operators (3) into matrix operators [46]. Thus, substituting the transformed operators to equation (7), we come to Hermitian eigenvalue problem parameterized with two-dimensional wave vector.

However, the additional difficulties of this computational scheme may be associated with full or partial mixing between computed eigenspectrum and the so-called "spurious" solutions [47] described below.

*2.3. The problem of "spurious" states*

Spurious solutions exhibiting unphysical behavior can be found in different parts of the simulated energy spectrum including the band gap. These states are characterized by incorrect subband bending in **k**-space, as well as high-frequency spatial oscillations. On the other hand, presence and features of these solutions strongly depend on the choice of numerical procedure [48].

In order to describe the reason for the appearance of such unphysical states let us consider a "toy" model $2\times 2$ **k·p**-Hamiltonian describing the interaction of nondegenerate conduction and valence bands:

$$\hat{H}_{2\times 2} = \begin{pmatrix} E_C + \tilde{A}k^2 & i\tilde{P}k \\ -i\tilde{P}k & E_V + \tilde{L}k^2 \end{pmatrix}, \quad (9)$$

where $E_C$, $E_V$ are the positions of the conduction and valence band edges respectively, $\tilde{P}$ is the momentum matrix element. The 8-band Kane Hamiltonian used in this work has blocks equivalent to (9), so the following discussion is applicable for the considered Hamiltonian.

The numerical solution of energy spectrum problem for an infinite discrete mesh with FDM or finite-elements-based scheme results in the eigenvalue problem:

$$\sum_j H_{i,j}^{(D)} \Psi_j = E \Psi_i, \quad (10)$$

where $i$, $j$ are the mesh node indices, and the discrete Hamiltonian operator $H_{i,j}^{(D)}$ has the three-diagonal form [49]:

$$H_{i,j}^{(D)} = \begin{pmatrix} \ddots & \vdots & \vdots & \vdots & \cdot \\ \cdots & \hat{\varepsilon} & \hat{t} & 0 & \cdots \\ \cdots & \hat{t} & \hat{\varepsilon} & \hat{t} & \cdots \\ \cdots & 0 & \hat{t} & \hat{\varepsilon} & \cdots \\ \cdot & \vdots & \vdots & \vdots & \ddots \end{pmatrix}, \quad (11)$$

where $\hat{\varepsilon}$ and $\hat{t}$ are the matrices describing the discrete form of the Hamiltonian. For the Hamiltonian (9) they are written as:

$$\hat{\varepsilon} = \begin{pmatrix} E_C + \frac{2\tilde{A}}{dz^2} & 0 \\ 0 & E_C + \frac{2\tilde{L}}{dz^2} \end{pmatrix}, \hat{t} = \begin{pmatrix} -\frac{\tilde{A}}{dz^2} & \frac{\tilde{P}}{2dz} \\ -\frac{\tilde{P}}{2dz} & -\frac{\tilde{L}}{dz^2} \end{pmatrix}. \quad (12)$$

By applying the Bloch theorem to (10) one yields the dispersion equation:

$$\left(\hat{\varepsilon} + \hat{t}\exp(ikdz) + \hat{t}^+ \exp(-ikdz)\right)\Psi = E(k)\Psi. \quad (13)$$

The solution $E(k)$ of the dispersion relation (13) has at least two extrema: the first one corresponding to the actual physical solution of the Hamiltonian (9) at $k = 0$ and the second one at $k = \pi/dz$ (unphysical valley) responsible for the presence of "spurious" states [50]. Then, substituting $k = \pi/dz$ to the dispersion equation (13) for FDM version of the equation (7) we obtain the energy position $E_{UV}$ of the unphysical valley which can be written as:

$$E_{UV} = E_C + \frac{4}{dz^2}\left(\frac{\hbar^2}{2m_0} + A\right), \quad (14)$$

where $m_0$ is the free electron mass, and $A$ describes quadratic-in-**k** interaction of the conduction band with the remote bands.

Decreasing of grid spacing results in shift of unphysical valley with respect to conduction band minimum. The direction of the shift is determined by the sign of $\hbar^2/2m_0 + A$ in (14). Neglecting the interaction with bands not included in the 8-band ***k·p***-Hamiltonian, we can make the shift to be positive by setting $A = 0$ according to Bastard [51]. In other words, to eliminate the problem of spurious states one has to fix the conduction band term of the Hamiltonian to be $\hbar^2 \mathbf{k}^2/2m_0$ and recalculate the interband momentum matrix term $P$ by means the electron effective mass value. Thus, adjusting the grid spacing we shift up the energy values of spurious solutions eliminating their mixing with physical states.

*2.4. Extracting the Dresselhaus and Rashba linear-in-k parameters from numerical calculations*

During numerical solution of equation (7) one get the electron energy dispersion:

$$E_{n,s}(k) = \frac{\hbar^2}{2m}\mathbf{k}^2 + \frac{s}{2}\Delta E_n(\mathbf{k}), \qquad (15)$$

where $\Delta E_n(\mathbf{k})$ is the spin splitting of $n^{\text{th}}$ electron subband. Angle dependence of this energy splitting can be written as

$$\Delta E_n(\mathbf{k}) = 2k\sqrt{\alpha_n^2 + \beta_n^2 + 2\alpha_n\beta_n \sin 2\varphi}, \qquad (16)$$

where $\varphi = \arctan(k_y/k_x)$.

For realization of the PSH state it is important to know the ratio between Rashba and Dresselhaus parameters. Our aim is to reveal a tendency in the behavior of their values in the single QWs with variable parameters. For direct extracting of the individual contribution $\alpha_n$ and $\beta_n$ values we input the linear-in-***k*** coefficients of the spin splitting $b_{n\pm}$ along corresponding direction $\Sigma_\pm = (1, \pm 1)$ in the momentum space:

$$\Delta E_n(k_x = k/\sqrt{2}, k_y = \pm k/\sqrt{2}) = 2|\alpha_n \pm \beta_n|k = b_{n\pm}k. \qquad (17)$$

By numerically determining coefficients $b_{n\pm}$ we get $\alpha_n$ and $\beta_n$ values from the equation (17) for each subband. The proposed approach is similar to one used earlier in [52] for experimental extraction of Rashba and Dresselhaus SOC parameters. It should be noted that Rashba and Dresselhaus contributions depend on a number of macroscopic characteristics, such as parameters of materials forming a QW, its width, doping profile and growth temperature, carrier density.

Thus, in this work we theoretically investigate the SOC in InGaAs/GaAs QW for different width $d$ with variable indium content using the 8-band envelope function approximation and FDM. We show that the behavior of the Dresselhaus and Rashba parameters varies greatly with a QW subband index. Besides we defined the QW parameters required for the PSH state formation.

The suggested method to calculate the spin splitting and the SOC parameters is self-sufficient and can be viewed as one of the main results of the work.

**3. Calculation results and discussion**

*3.1. Dresselhaus parameter in the symmetric quantum well of different width*

Let us consider first a single symmetric $\text{In}_x\text{Ga}_{1-x}\text{As}/\text{GaAs}$ QW grown along the direction $[001]$. Kane Hamiltonian parameters for $\text{In}_x\text{Ga}_{1-x}\text{As}$ semiconductor alloys were taken from Ref. [45]. Because in symmetric structures, due to parity, the intrasubband SO interaction contains only the Dresselhaus contribution, therefore we calculate this parameter for each subband with different QW width $d$.

In the simplest approximation of infinitely deep well the Dresselhaus parameter has following form (see, e.g., [53]):

$$\beta_n = \frac{\pi^2 n^2 \gamma_D}{d^2}. \qquad (18)$$

According to this approximation the Dresselhaus term monotonically decreases with QW width for all quantum subbands.

If we solve this problem numerically, by means of method described in 2.4, we get a different result for the Dresselhaus parameter with increasing width QW. Figure 1 demonstrates the dependence of the Dresselhaus parameter absolute value on QW width calculated for three lower QW subbands. It is evident that only for the first subband this parameter behaves according to the equation (18), i.e. monotonically drops with increasing QW width. As the subband index growth, the dependence $\beta_n(d)$ becomes complicated, moreover, for the third subband we observe a change in the sign of the Dresselhaus parameter at a certain value of the well width.

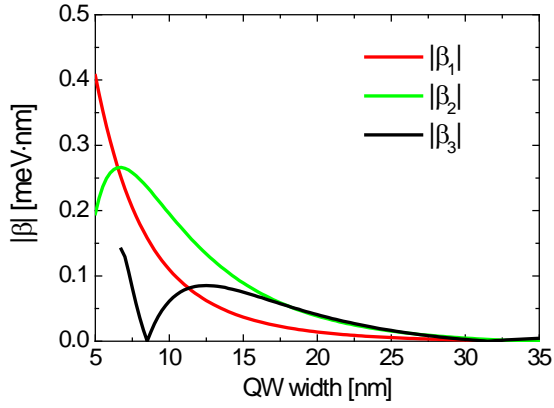

**Figure 1.** Absolute value of the Dresselhaus parameter for $In_xGa_{1-x}As/GaAs$ ($x = 0.4$) QWs, calculated for three electron subbands as a function of QW width.

One reason of such a behaviour is incorrectness of equation (18) for finite-depth QWs. In reality the electron wave function is not completely localized inside the QW but penetrates into barriers layers [53]. Moreover, the fraction of this penetration depends significantly on the heterostructure materials parameters and system's geometry.

*3.2. Dresselhaus parameter in the symmetric quantum well with variable In content*

In this section we suggest an explanation of the non-trivial behavior of the Dresselhaus parameter $\beta_n(d)$ for different subband index, QW width and indium content. According to equation (5) $\beta_n(d)$ depends on Dresselhaus constants both in well and barrier regions. Difference in the well and barrier materials leads to the electron density redistribution between them and, as a consequence, changing of contributions to the Dresselhaus parameter. It is evident that the resultant Dresselhaus parameter should be proportional to the wave function fraction multiplied by corresponding bulk Dresselhaus constant $\gamma_D(z)$ both of the well and the barrier material. In other words, when calculating parameter $\beta_n$, one need to take into account the real value $\gamma_D(z)$ for each coordinate point of the structure. As known for $In_xGa_{1-x}As$ solid solutions, the bulk Dresselhaus constant $\gamma_D$ varies drastically with indium content, hence it can differently contribute to resultant value of the Dresselhaus parameter and lead not only to the value changing but also to the sign inversion [54]. Consequently, the Dresselhaus parameter can unusually behave with simultaneous electron subband index and QW width increasing [55].

The indium content varying results in changing of band offsets at the interface, therefore the effective depth of the well changes for all subbands. However, behavior of the Dresselhaus parameter for the third subband looks the most intriguing (figure 1), so let us consider it in more detail. At figure 2 there are shown the dependence of the Dresselhaus parameter modulus $|\beta_3|$ on the QW width for several values of In content. The dependence is nonmonotonic with maximum for QW width $\sim 12$ nm. Moreover, the Dresselhaus parameter substantially decreases with increase of In content $x$. Such a behaviour is due to change of $\gamma_D$ in the well region, including its sign. Besides, the linear-in-*k* Dresselhaus parameter equals zero in QWs with $x > 0.36$ at some certain QW width.

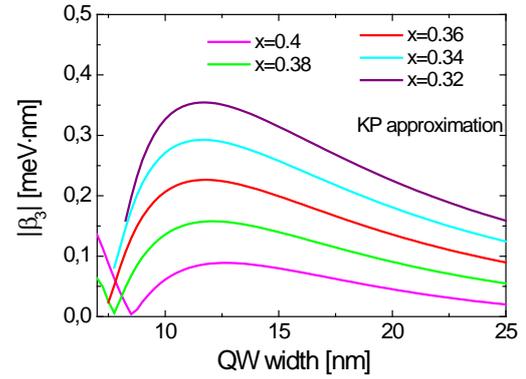

**Figure 2.** Absolute value of the Dresselhaus parameter calculated for third electron subband ($\beta_3$) as a function of QW width with variable In content $x$.

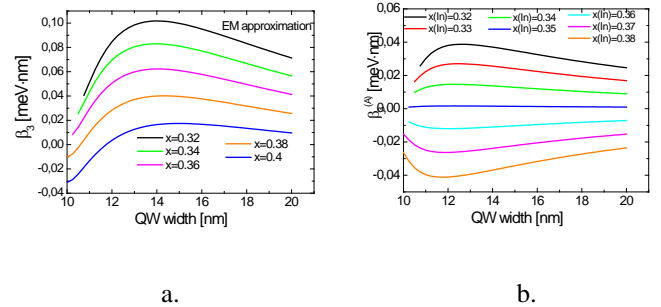

a.				b.

**Figure 3.** **(a)** The well region component $\beta_3^{(A)}$ of the Dresselhaus parameter for third electron subband calculated using single-band effective mass aprproximation. **(b)** Total (sum of the well and barrier regions) value $\beta_3$ as a function of QW width for third electron subband. $x$ stands for In content in well layer.

In order to estimate the Dresselhaus contribution for each regions (well and barrier), we split the envelope function into two fractions: corresponding to the well layer region (A) and barrier region (B). For symmetric QW it is sufficient to make calculations in the framework of the effective mass approximation [56]: such an approach gives qualitative description. The Dresselhaus parameter $\beta_n(d)$ for the $n^{th}$ subband can be presented as a sum of two contributions:

$$\beta_n = \beta_n^{(A)} + \beta_n^{(B)},$$
$$\beta_n^{(A,B)} = \gamma_D^{(A,B)} \int_{A,B} dz\, \psi_n^+(z) \frac{\partial^2}{\partial z^2} \psi_n(z), \quad (19)$$

where $\beta_n^{(A)}$ and $\beta_n^{(B)}$ correspond to $In_xGa_{1-x}As$ well and GaAs barrier regions, respectively, $\psi_n(z)$ is the envelope function obtained in the framework of the single-band effective mass method. According to our estimates, the quantitative error of the effective mass method in comparison with the Kane model (or, in other words, the error of the averaging procedure based on the effective mass approximation described in detail in Sec. 2) for $In_{0.4}Ga_{0.6}As/GaAs$ QW is about 1.5 times for the first electron subband and about 3 times for the third subband.

Figure 3(a) shows the Dresselhaus parameter $\beta_3^{(A)}(d)$ of the A region and figure 3(b) demonstrates the total value $\beta_3(d)$ as function of QW width for the third subband with different In content. Here we are taking into account that the penetration of the wave function into the barrier regions varies both with the well width and with subband index. Obviously the third subband is characterized by more significant electron density redistribution and wave function penetration into the barrier. It is seen that A-region contribution is nonmonotonic with QW width and increasing for any indium content. We believe that it occurs due to the simultaneous superposition of two physical effects: the barrier wave function part (B) is increasing but the quantum state average kinetic energy is decreasing. In other words, as the subband index increases, the kinetic energy becomes smaller, but the penetration into the barrier increases.

Summing the two separately calculated well/barrier components according to (19) we obtain $\beta_3(d)$ dependence, which is also nonmonotonic. Moreover, in InGaAs the bulk Dresselhaus constant changes its sign from some indium content. It is seen that this dependence (figure 3(b)) qualitatively coincides with the behavior of the parameter $\beta_3$ (at the corresponding In content) calculated in the framework of the 8-band $\mathbf{k}\cdot\mathbf{p}$-model (figure 2). Thus, the results of multi-band calculations and effective mass method show, that for higher QW subbands the averaged Dresselhaus parameter can vary within wide range, turning to zero at the definite combinations of QW parameters.

It should be noted that obtained result is specific for peculiar QW materials $In_xGa_{1-x}As/GaAs$ because bulk Dresselhaus constant $\gamma_D$ strongly varies from well to barrier. In particular, the same Dresselhaus parameter behavior is unreachable for the heterostructures based on GaAs/GaAsSb materials because bulk Dresselhaus constant $\gamma_D$ has the same sign at any solid solution contents.

*3.3. PSH at the asymmetric QW potential, caused by the transverse electric field*

Further we consider asymmetric $In_xGa_{1-x}As/GaAs$ QW potential profile with different width, in which both Dresselhaus and Rashba SOC are present. Asymmetric potential is obtained by means of an external electric field ($E = 0 \div 0.4$ mV/nm) applied in the structure growth direction. Solving the equation (7), we calculate the energy spectrum taking into account SOC, described by superposition of Rashba and Dresselhaus effects. In this Section our aim is to reveal the electric filed range that satisfy the conditions for the PSH regime realization [17, 18] with different QW width and In content. It is known the PSH state corresponds to the numerical equality of Rashba and Dresselhaus constants ($|\alpha_n| = |\beta_n|$) (see figure 4).

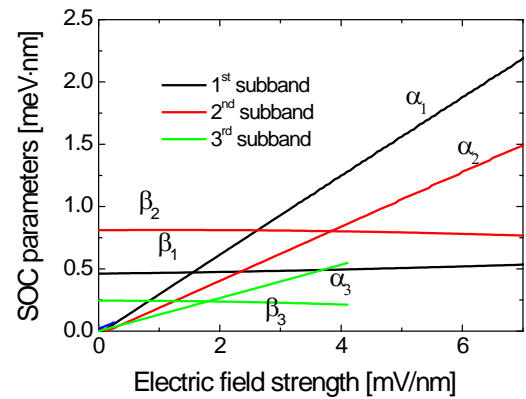

**Figure 4.** Rashba and Dresselhaus parameters ($\alpha_n$ and $\beta_n$) dependence on the electric field strength for three lower electron subbands in $In_xGa_{1-x}As/GaAs$ QW with the In content $x = 0.4$ and well width $d = 10$ nm.

By means the method, described above, we calculate SOC parameters in $In_xGa_{1-x}As/GaAs$ QW structure for all subbands as a function of applied transverse electric field (figure 4).

It is obvious that PSH state for the first subband can be realized with a small In content in the QW ($x$ is about 0.2). At the same time it is not difficult to show that increasing of indium content in QW should lead to possibility of PSH formation for excited subbands.

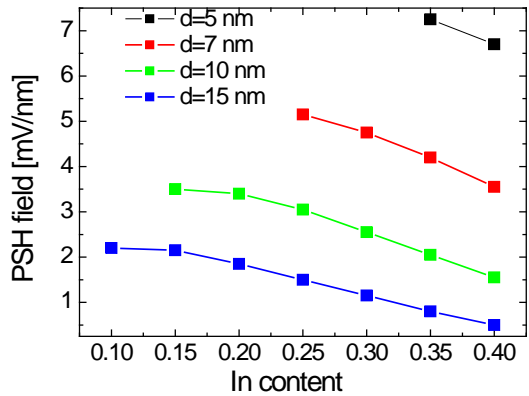

**Figure 5.** PSH field values for the first electron subband ($|\alpha_1|=|\beta_1|$) as a function of indium content in In$_x$Ga$_{1-x}$As/GaAs QW with different QW width $d=5,7,10,15$ nm.

Thus, for QW with a large In content ($x$ is about 0.4), applying an electric field one can obtain a PSH regime with various combinations of heterostructure geometry and In content. At figure 4 it is depicted the Rashba and Dresselhaus parameters crossing points $|\alpha_n|=|\beta_n|$, corresponding to PSH states for three electron subbands. One can see from this figure, that the parameter $\alpha_n$ induced by asymmetry structure is almost linear function of the external field and is almost independent on the width well [57]. The Dresselhaus parameter, in its turn, is not vary with the increasing field and is determined, in principal, by subband index, In content and QW width (Sec. 3.1). Since the Rashba parameter is well controlled by the electrical field (figure 4), we can draw a conclusion, that for PSH realization one can adjust it by gate voltage for any Dresselhaus parameter determined by the structure itself.

Also, we identified that the Dresselhaus parameter nonmonotonic dependence on the QW width with the subband index increasing results in a nontrivial behavior of the PSH field (the electric field that allows us to realize the PSH condition). From figure 4 it is seen, that the PSH field required for the third subband (red crossing point) has lesser value than it for second subband (green crossing point) at a given quantum well width.

Alternating Dresselhaus parameter behavior for the third subband is very interesting result from fundamental point of view. But, *senso stricto*, experimentally observed QW electronic properties are mostly determined by the first subband, which is typically populated with majority of carriers. Therefore, we limit further our considerations to the simulation of QW electronic properties for only one subband. In particular, one of the aims of this research is to show that even for the first subband we can create the PSH regime in the wide range of QW material and geometric parameters at the real electric field strength.

Smoothly varying the indium content and QW width let us observe how electric field the required for PSH symmetry point $|\alpha_1|=|\beta_1|$ is changing (see figure 5). Figure 5 shows that the PSH field almost linearly decreases with In content increasing (the 10% shift of In content leads to the change of PSH field around 0.5 mV/nm). Note also, that for all considered QWs the relation between the PSH field and In content depends weakly on the QW width. It is also shown that for fixed indium content, PSH field decreases rapidly with QW width growth.

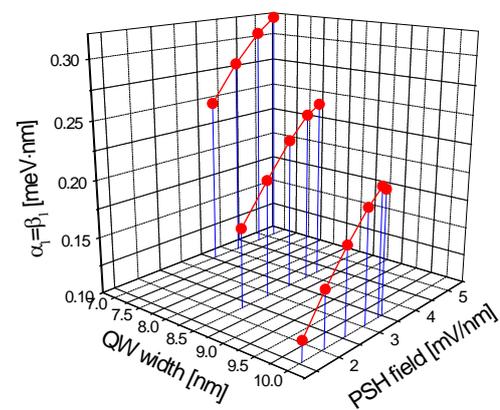

**Figure 6**. PSH SOC parameters, calculated for the first electron subband ($|\alpha_1|=|\beta_1|$) as a function of electric field strength in In$_x$Ga$_{1-x}$As/GaAs QW with variable indium content ($x=0.1...0.4$) and QW width $d=7,8.5,10$ nm.

Figure 6 demonstrates a summary of our calculations for asymmetric QWs. On this figure one can see the first subband SOC parameter (the Rashba parameter equals the Dresselhaus one) as a function of the PSH field for different In contents. Thus, by combining the QW width and In content varying we can realize the PSH regime with different precession frequency. It is necessary to emphasize that calculated PSH fields can be easily obtained experimentally, that is important for possible applications.

**4. Conclusion**

By direct numerical calculations, a technique has been proposed that makes it possible to extract the individual contribution of the Dresselhaus and Rashba SOC parameters for the QWs conduction band of based on III-V semiconductors with a zinc blende structure in the presence of a transverse uniform electric field.

The calculations carried out in this paper show that the asymmetry of the potential contributes to an increase in the spin splitting for all electron subbands in the QW. The Rashba parameter calculated in the framework of the 8-band Kane model depends linearly on the applied electric field, remaining practically insensitive to the width of the QW and the index of the size-quantized subband. In turn, the Dresselhaus parameter is practically independent on the external applied field, but it varies significantly with the subband index and with the well width.

Taking into account the redistribution of the electron wave function between the well and the barrier regions in the symmetric finite-depth QW, we explain the nonmonotonic Dresselhaus parameter dependence on the electron subband index and, in particular, its alternating behavior with QW width for third subband. Moreover, as a specific result we obtain the possibility to eliminate the linear-in-$k$ Dresselhaus term in QW at the certain parameters.

We have also demonstrated that applying of external transverse electric field to $In_xGa_{1-x}As/GaAs$ QW allows one to effectively control the ratio between the Rashba and Dresselhaus parameters. The requirement of $|\alpha_n|=|\beta_n|$ can be fulfilled by the variation of both Rashba and Dresselhaus terms, which in turn depend on a variety of parameters, such as subband index, electric field strength (external gate voltage), QW materials and structural design. Thus, by combining the configuration of the structure and the external electric field, one can realize the PSH states.

It should be noted both SOC parameters are highly sensitive to the heterostructure material kind. We showed earlier [58] that for another pair of heteromaterials (for example InGaSb/GaSb), the equality of the Rashba and Dresselhaus constants is extremely difficult with available electric fields.

Our results can be used for design of semiconductor spintronics structures with required characteristics. Experimental observation of predicted spin splittings is allowed at temperatures around 4 K, so our calculations can be used for interpretation of experimental results on transport and optical properties of heterostructures with strong SOC.

**Acknowledgements**

The work was supported by the Lobachevsky University development program under the Russian academic excellence project 5-100.